\documentclass[12pt]{article}
\usepackage{xspace} 
\usepackage{graphicx}

\def\CP                {\ensuremath{C\!P}\xspace}
\def\CPT               {\ensuremath{C\!PT}\xspace}
                 
\def\jpsi     {\ensuremath{{J\mskip -3mu/\mskip -2mu\psi\mskip 2mu}}\xspace}
\def\bquark    {\ensuremath{b}\xspace}
\def\squark    {\ensuremath{s}\xspace}

\def\pion  {\ensuremath{\pi}\xspace}
\def\piz   {\ensuremath{\pion^0}\xspace}

\def\pip   {\ensuremath{\pion^+}\xspace}
\def\pim   {\ensuremath{\pion^-}\xspace}

\def\pipm  {\ensuremath{\pion^\pm}\xspace}
\def\pimp  {\ensuremath{\pion^\mp}\xspace}
\def\kaon  {\ensuremath{K}\xspace}
\def\Kp    {\ensuremath{\kaon^+}\xspace}
\def\Km    {\ensuremath{\kaon^-}\xspace}
\def\Kpm   {\ensuremath{\kaon^\pm}\xspace}

\def\Kstarz  {\ensuremath{\kaon^{*0}}\xspace}
\def\Dz      {\ensuremath{D^0}\xspace}

\def\Dstarp  {\ensuremath{D^{*+}}\xspace}

\def\Bu      {\ensuremath{B^+}\xspace}
\def\Bub     {\ensuremath{B^-}\xspace}
\def\Bp      {\ensuremath{\Bu}\xspace}
\def\Bm      {\ensuremath{\Bub}\xspace}
\def\Bpm     {\ensuremath{B^\pm}\xspace}

\def\Bd      {\ensuremath{B^0}\xspace}
\def\Bs      {\ensuremath{B^0_\squark}\xspace}
\newcommand{\tev}{\ifthenelse{\boolean{inbibliography}}{\ensuremath{~T\kern -0.05em eV}\xspace}{\ensuremath{\mathrm{\,Te\kern -0.1em V}}\xspace}}
\newcommand{\gev}{\ensuremath{\mathrm{\,Ge\kern -0.1em V}}\xspace}
\newcommand{\mev}{\ensuremath{\mathrm{\,Me\kern -0.1em V}}\xspace}
\newcommand{\kev}{\ensuremath{\mathrm{\,ke\kern -0.1em V}}\xspace}
\newcommand{\ev}{\ensuremath{\mathrm{\,e\kern -0.1em V}}\xspace}
\newcommand{\gevc}{\ensuremath{{\mathrm{\,Ge\kern -0.1em V\!/}c}}\xspace}
\newcommand{\mevc}{\ensuremath{{\mathrm{\,Me\kern -0.1em V\!/}c}}\xspace}
\newcommand{\gevcc}{\ensuremath{{\mathrm{\,Ge\kern -0.1em V\!/}c^2}}\xspace}
\newcommand{\gevgevcccc}{\ensuremath{{\mathrm{\,Ge\kern -0.1em V^2\!/}c^4}}\xspace}
\newcommand{\mevcc}{\ensuremath{{\mathrm{\,Me\kern -0.1em V\!/}c^2}}\xspace}

\def\invfb   {\ensuremath{\mbox{\,fb}^{-1}}\xspace}
\newcommand{\stat}{\ensuremath{\mathrm{\,(stat)}}\xspace}
\newcommand{\syst}{\ensuremath{\mathrm{\,(syst)}}\xspace}

\def\pipipi {\ensuremath{{\Bpm \to \pip \pim \pipm}}\xspace}
\def\kpipi {\ensuremath{{\Bpm \to \Kpm \pip \pim}}\xspace}
\def\kkpi {\ensuremath{{\Bpm \to \Kp \Km \pipm}}\xspace}
\def\kkk {\ensuremath{{\Bpm \to \Kpm \Kp \Km}}\xspace}

\def\ppk {\ensuremath{{\Bpm \to p \bar{p} K^{\pm}}}\xspace}
\def\jpsik {\ensuremath{{\Bpm \to \jpsi \Kpm}}\xspace}

\def\jpsiks {\ensuremath{{\jpsi \Kpm}}\xspace}
\def\mmkpi {\ensuremath{m^2_{\Kpm \pimp}}\xspace}                    
                       
\def\mmpipi {\ensuremath{m^2_{\pip \pim}}\xspace}                       
\def\mmkk {\ensuremath{m^2_{\Kp \Km}}\xspace}

\def\mmkklow {\ensuremath{m^2_{\Kp \Km\,{\rm low}}}\xspace}   
\def\mmkkhi {\ensuremath{m^2_{\Kp \Km\,{\rm high}}}\xspace}     
\def\mmpipilow {\ensuremath{m^2_{\pip \pim\,{\rm low}}}\xspace} 
\def\mmpipihi {\ensuremath{m^2_{\pip \pim\,{\rm high}}}\xspace}  
\def\acp {\ensuremath{A_{\CP}}\xspace}
\def\acpraw {\ensuremath{A_{\rm raw}}\xspace}

\def\adet {\ensuremath{A_{\rm D}}\xspace}

\def\Kbar  {\kern 0.2em\overline{\kern -0.2em K}{}\xspace}
\def\Kstarz  {\ensuremath{\kaon^{*0}}\xspace}
\def\Kstarzb {\ensuremath{\Kbar^{*0}}\xspace}
\def\ppk {\ensuremath{\Bpm \to p\bar{p}\Kpm}\xspace}
\def\phik {\ensuremath{\Bs \to \phi\Kstarzb}\xspace}
\def\kkpiz {\ensuremath{\Bd \to \Kp\Km \piz}\xspace}
\def\bdkpi {\ensuremath{\Bd \to \Kp\pim}\xspace}
\def\bskpi {\ensuremath{\Bs \to \Km\pip}\xspace}
\def\bds {\ensuremath{B_{(\squark)}^0 }\xspace}
\def\splot{\ensuremath{_s{\cal P}lot}\xspace}
\def\KS    {\ensuremath{\kaon^0_{\rm\scriptscriptstyle S}}\xspace} 

\newcommand\pubnumber{LHCb-PROC-2013-046}
\newcommand\pubdate{\today}

\def\cbpf{Centro Brasileiro de Pesquisas F\'isicas\\
Rio de Janeiro, Brazil}
\def\support{\footnote{On behalf of the LHCb collaboration.}}

\def\Title#1{\begin{center} {\Large #1 } \end{center}}
\def\Author#1{\begin{center}{ \sc #1} \end{center}}
\def\Address#1{\begin{center}{ \it #1} \end{center}}

\newcommand\pubblock{\rightline{\begin{tabular}{l} \pubnumber\\
         \pubdate  \end{tabular}}}
\newenvironment{Abstract}{\begin{quotation}  }{\end{quotation}}
\newenvironment{Presented}{\begin{quotation} \begin{center} 
             PRESENTED AT\end{center}\bigskip 
      \begin{center}\begin{large}}{\end{large}\end{center} \end{quotation}}
\def\Acknowledgements{\bigskip  \bigskip \begin{center} \begin{large}
             \bf ACKNOWLEDGEMENTS \end{large}\end{center}}

\def\beq{\begin{equation}}
\def\eeq#1{\label{#1}\end{equation}}
\def\eeqn{\end{equation}}

\def\beqa{\begin{eqnarray}}
\def\eeqa#1{\label{#1}\end{eqnarray}}
\def\eeqan{\end{eqnarray}}

\let\bar=\overbar

\def\Dslash{\not{\hbox{\kern-4pt $D$}}}
\def\dslash{\not{\hbox{\kern-2pt $\del$}}}

\def\msb{{\bar{\ssstyle M \kern -1pt S}}}

\begin{document}
\begin{titlepage}
\pubblock

\vfill
\Title{Studies of charmless $B$ decays including $C\!P$ violation effects}
\vfill
\Author{Irina Nasteva\support}
\Address{\cbpf}
\vfill
\begin{Abstract}
The latest experimental results in charmless $B$ decays are presented with a focus on $C\!P$ violation measurements. These include the first observation of $C\!P$ violation in $B^0_s$ decays, evidence for $C\!P$ violation in charmless three-body $B^{\pm}$ decays, branching fraction measurements of $B^{\pm} \rightarrow p\bar{p}K^{\pm}$ decays and the first observation of the decay $B^0_s \rightarrow \phi \bar{K}^{*0}$ from LHCb, a comparison of $B^{\pm}\rightarrow K^+K^-K^{\pm}$ $C\!P$ violation measurements between LHCb and BaBar, and the first evidence for the decay $B^0 \rightarrow K^+K^- \pi^0$ obtained by Belle.
\end{Abstract}
\vfill
\begin{Presented}
Flavor Physics and CP Violation 2013 (FPCP--2013)\\
Buzios, Rio de Janeiro, Brazil, May 19--24, 2013
\end{Presented}
\vfill
\end{titlepage}
\def\thefootnote{\fnsymbol{footnote}}
\setcounter{footnote}{0}

\section{Introduction}

Charmless $B$ decays occur through loop diagrams and are thus sensitive to possible new physics contributions in the loops. 
Two- and three-body charmless $B$ decays proceed via \CP-conserving penguin diagrams and \CP-violating tree diagrams with $b\to u$ transitions, which give access to the $\gamma $ CKM angle. 
The tree-penguin interference allows to look for direct \CP violation in $B^0_{(\squark)}\to hh'$ and $\Bpm \to hhh'$ decays, where $h$ and $h'$ stand for hadrons. 
The direct \CP asymmetry of $B$ decays to a final state $f$ is defined as
\begin{equation}
\acp(B\to f) = \frac{ \Gamma(\bar{B}\to\bar{f}) - \Gamma(B \to f) }{ \Gamma(\bar{B}\to\bar{f}) + \Gamma(B \to f)} \, .
\label{eq:arawGamma}
\end{equation}

With the LHCb experiment~\cite{LHCbdet} currently running, and BaBar~\cite{BaBardet} and Belle~\cite{Belledet} analysing their full datasets, we present several new and recent results of \CP violation and branching fraction measurements of charmless $B$ decays. 
The LHCb results are obtained from $1\invfb$ of data from $pp$ collisions at a centre-of-mass energy of $\sqrt{s}=7$~TeV. 
Section~\ref{Bhh} presents the LHCb measurements of \CP violation in charmless two-body $B^0_{(\squark)}$ decays.
Section~\ref{Bhhh} details the results of studies of charmless three-body decays, including \CP violation measurements of \kpipi, \kkk, \kkpi and \pipipi from LHCb, a comparison of the \kkk results with BaBar, evidence for a new decay channel \kkpiz from Belle, and branching fraction measurements of \ppk decay components from LHCb. 
Finally, Section~\ref{BVV} describes the observation of the decay to two vector mesons \phik by LHCb.

\section{Charmless two-body $\bds$ decays}
\label{Bhh}

\begin{figure}[tb]
\centering
\includegraphics[width=0.8\linewidth]{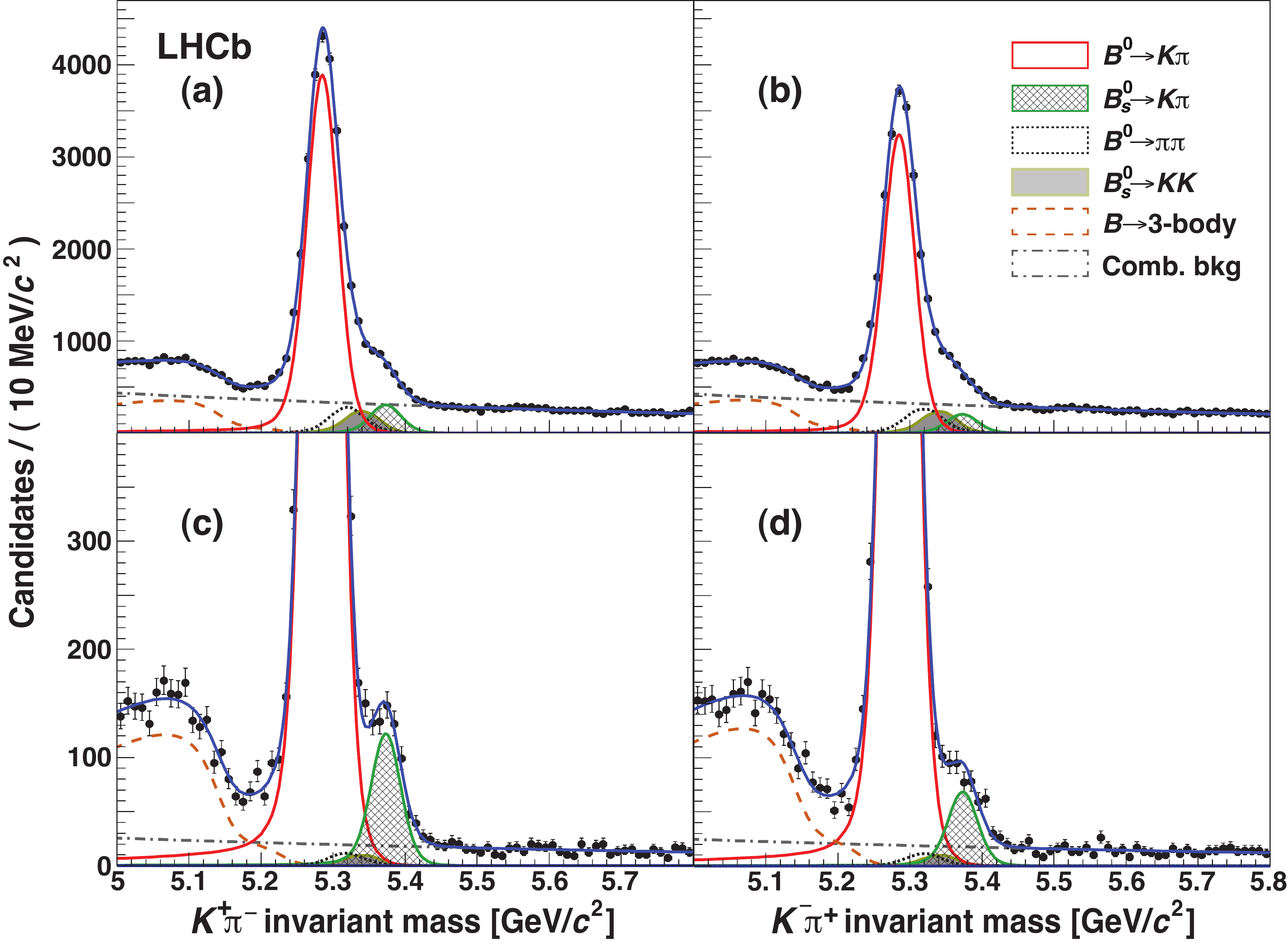}
\caption{Invariant mass spectra using the event selection optimised for (a, b) $\acp(\bdkpi)$ and (c, d) $\acp(\bskpi)$. Panels (a) and (c) show the $\Kp\pim$ invariant mass and panels (b) and (d) show the $\Km\pip$ invariant mass. The results of the unbinned maximum likelihood fits are overlaid.
}
\label{fig:B2Kpi}
\end{figure}

The LHCb experiment has recently measured the direct \CP-violating asymmetries of the decays \bdkpi and \bskpi with an unprecedented precision~\cite{BsCPV}.
The data samples of two-body decays are selected using two different sets of criteria aimed at obtaining the best sensitivity to \acp for \Bd and \Bs decays, respectively. 
The \Bs measurement requires a stronger rejection of combinatorial backgrounds, due to the lower probability for a \bquark quark to form a \Bs meson. 
After the kinematic and trigger selections, the two data samples are subdivided into different final states, using particle identification (PID) information~\cite{RICH}. 
The PID selection criteria applied to \Bs candidates are tighter than those for the \Bd sample. 
In order to determine the background contributions with one or more misidentified final-state particles (cross-feed backgrounds), the PID efficiencies and misidentification rates are determined from large samples of calibration data, containing $\Dstarp \to \Dz(\Km\pip)\pip$ and $\Lambda \to p \pim$ decays. 

The signal yields and raw asymmetries are obtained form unbinned extended maximum likelihood fits to the invariant mass spectra of the selected candidates, shown in Fig.~\ref{fig:B2Kpi}.  
The raw asymmetries are corrected for detection asymmetry effects due to detector reconstruction, acceptance and final-state particle interaction with matter, and for the $\bds-\bar{B}_{(\squark)}^0$ production asymmetry. 
The \CP asymmetry is calculated as
\begin{equation}
 \acp = \acpraw - A_{\Delta} \, ,
 \label{acpeq}
\end{equation}
where the correction $A_{\Delta}$ is the sum of the detection asymmetry $\adet(K\pi)$ and the production asymmetry $A_{\rm P}(\bds)$:
\begin{equation}
 A_{\Delta} = \zeta_{d(s)}\adet(K\pi)  + \kappa_{d(s)}A_{\rm P}(\bds)  \, 
\end{equation}
with $\zeta_d=1$, $\zeta_s=-1$ and $\kappa_{d(s)}$ are dilution factors due to mixing.

The detection asymmetries are determined from data using $\Dstarp \to \Dz(\Km\pip)\pip$ and $\Dstarp \to \Dz(\Kp\Km)\pip$ decays. 
From their time-integrated asymmetries the values $\adet(K\pi)=(-1.15\pm 0.23)\%$ for \bdkpi and $\adet(K\pi)=(-1.22\pm 0.21)\%$ for \bskpi decays are measured. 
The production asymmetries are taken from a time-dependent fit to the decay rate and asymmetry of the \Bd and \Bs samples, obtaining $A_{\rm P}(\Bd)=(0.1\pm 1.0)\%$ and $A_{\rm P}(\Bs)=(4\pm8)\%$. 
Finally, the \CP asymmetries of $\bds \to K\pi$ decays are measured to be
\begin{eqnarray}
\acp(\bdkpi)&=&-0.080 \pm 0.007\stat \pm 0.003\syst \, , \nonumber \\
\acp(\bskpi)&=&0.27 \pm 0.04\stat \pm 0.01\syst \, . \nonumber
\end{eqnarray}
The systematic uncertainties account for the PID calibration, the signal and background fit models, and the detection charge asymmetries. 
The significances of the results, calculated by dividing the central values by the sum in quadrature of the statistical and systematic uncertainties, are $10.5\sigma$ and $6.5\sigma$, respectively. 
The former is the most precise measurement of $\acp(\bdkpi)$, and the latter represents the first observation of direct \CP violation in the \Bs system.

\section{Charmless three-body $B$ decays}
\label{Bhhh}

\subsection{\CP violation in \kpipi and \kkk decays}
\label{secKpipi}

\begin{figure}[tb]
\centering
\includegraphics[width=0.48\linewidth]{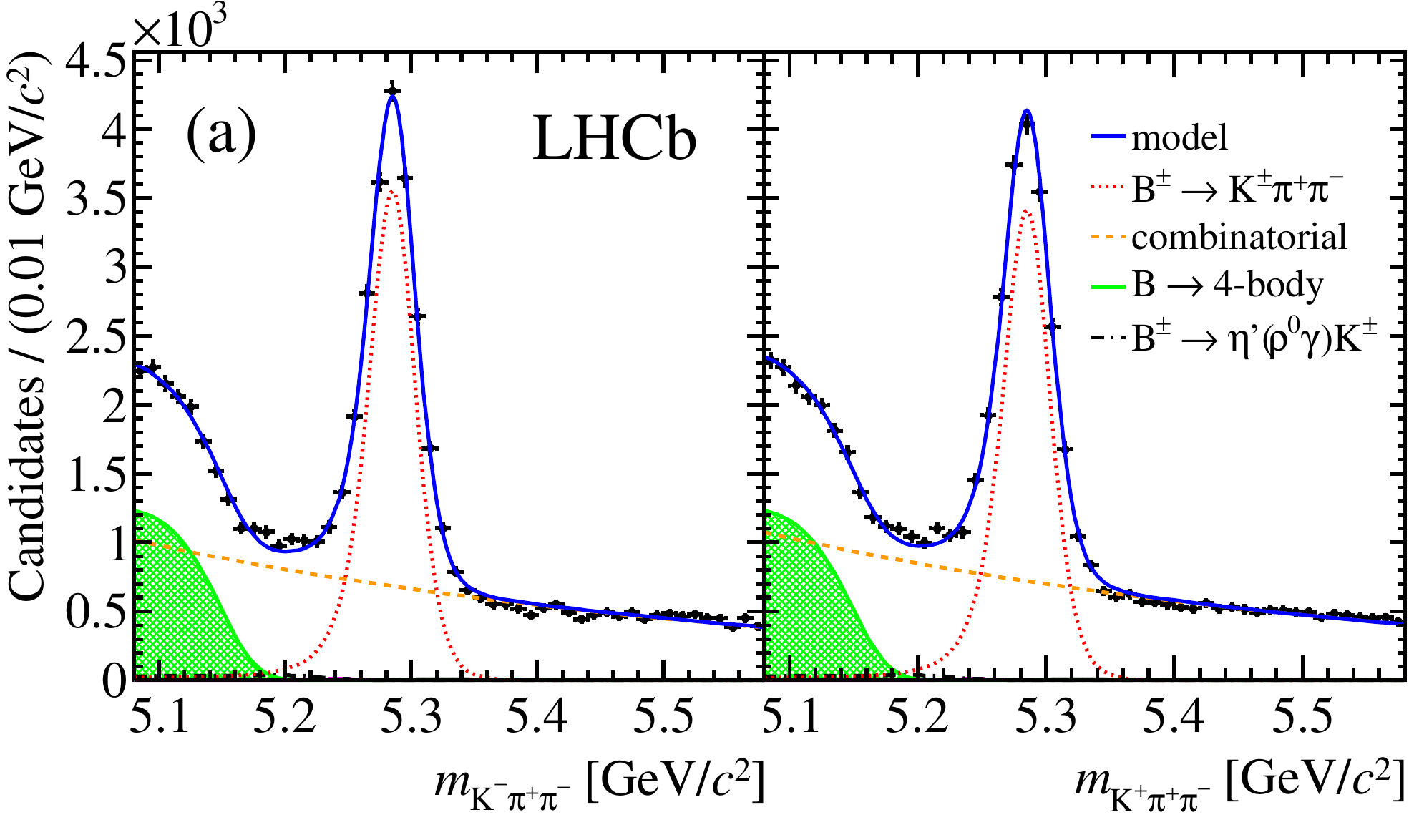}
\includegraphics[width=0.48\linewidth]{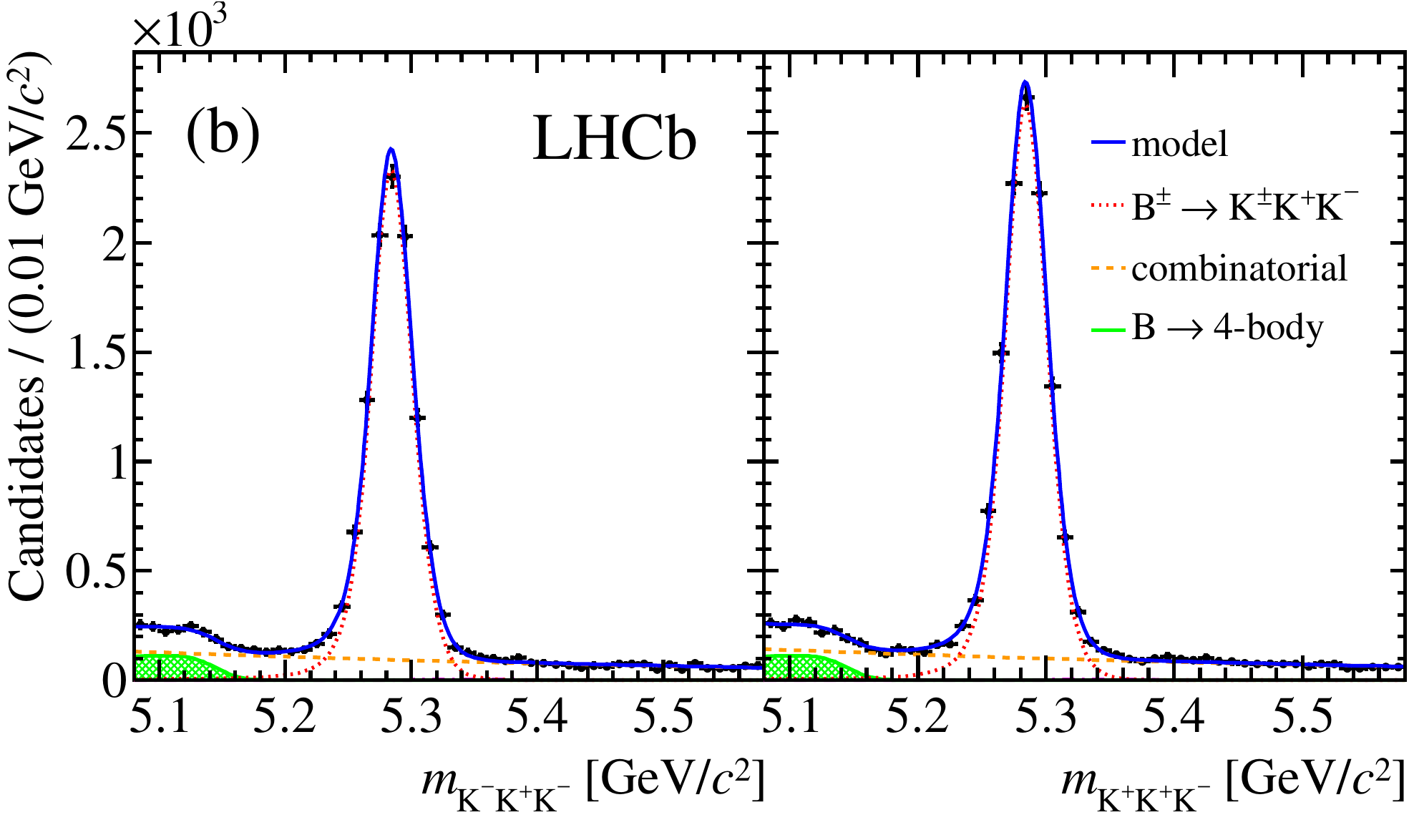}
\caption{Invariant mass spectra of (a) \kpipi and (b) \kkk decays. The left (right) panels show the \Bm (\Bp) modes. The results of the unbinned maximum likelihood fits are overlaid.
}
\label{fig:Kpipi}
\end{figure}

Decays of the type $\Bpm\to h^+h^-h'^{\pm}$ ($h,h'=K,\pi$) offer the possibility to study \CP violation patterns in the Dalitz plane in addition to the global asymmetries. 
\CP violation can appear through the penguin-tree interference, through intermediate resonant states or through final-state hadron rescattering~\cite{Marshak,Wolfenstein,Branco,Bigi} between two or more channels with the same flavour quantum numbers, such as \kpipi and \kkk.
The latter effect, also called ``compound \CP violation''~\cite{Soni2005}, is constrained by \CPT invariance so that the sum of all partial decay widths of a particle, to channels with the same final-state quantum numbers, must match those of its antiparticle. 
The joint study of \kpipi and \kkk decays was motivated by their \CPT connection.

The LHCb collaboration presented for the first time a measurement of the direct \CP asymmetry in \kpipi and \kkk decays with unprecedented precision~\cite{LHCb-PAPER-2013-027}. 
The data sample is defined by a common kinematic selection of \Bpm decays to three charged mesons, followed by a PID selection to distinguish different final states. 
The raw asymmetries are obtained from fits to the invariant mass spectra, shown in Fig.~\ref{fig:Kpipi}, and are corrected for the variation of the acceptance efficiency in phase space. 
The inclusive \CP asymmetries are then determined using Eq.~(\ref{acpeq}), where the correction term $A_\Delta$ is found from a data control sample of \jpsik decays. 
The inclusive \CP asymmetries are calculated separately according to different trigger selections and then averaged to obtain
\begin{eqnarray}
\acp(\kpipi)&=& 0.032\pm 0.008\stat \pm 0.004\syst \pm 0.007(\jpsiks) \, , \nonumber \\
\acp(\kkk)&=& -0.043 \pm 0.009\stat \pm 0.003\syst  \pm 0.007(\jpsiks)\, , \nonumber
\end{eqnarray}
where the systematic uncertainties account for the signal and background fit models, trigger asymmetry and acceptance correction, and the last uncertainty is due to the \CP asymmetry of the \jpsik control channel~\cite{PDG}. 
The significances of the inclusive charge asymmetries are $2.8\sigma$ and $3.7\sigma$, respectively. The latter represents the first evidence for an inclusive charge asymmetry in charmless three-body $B$ decays.

\begin{figure}[tb]
\centering
\includegraphics[width=0.48\linewidth]{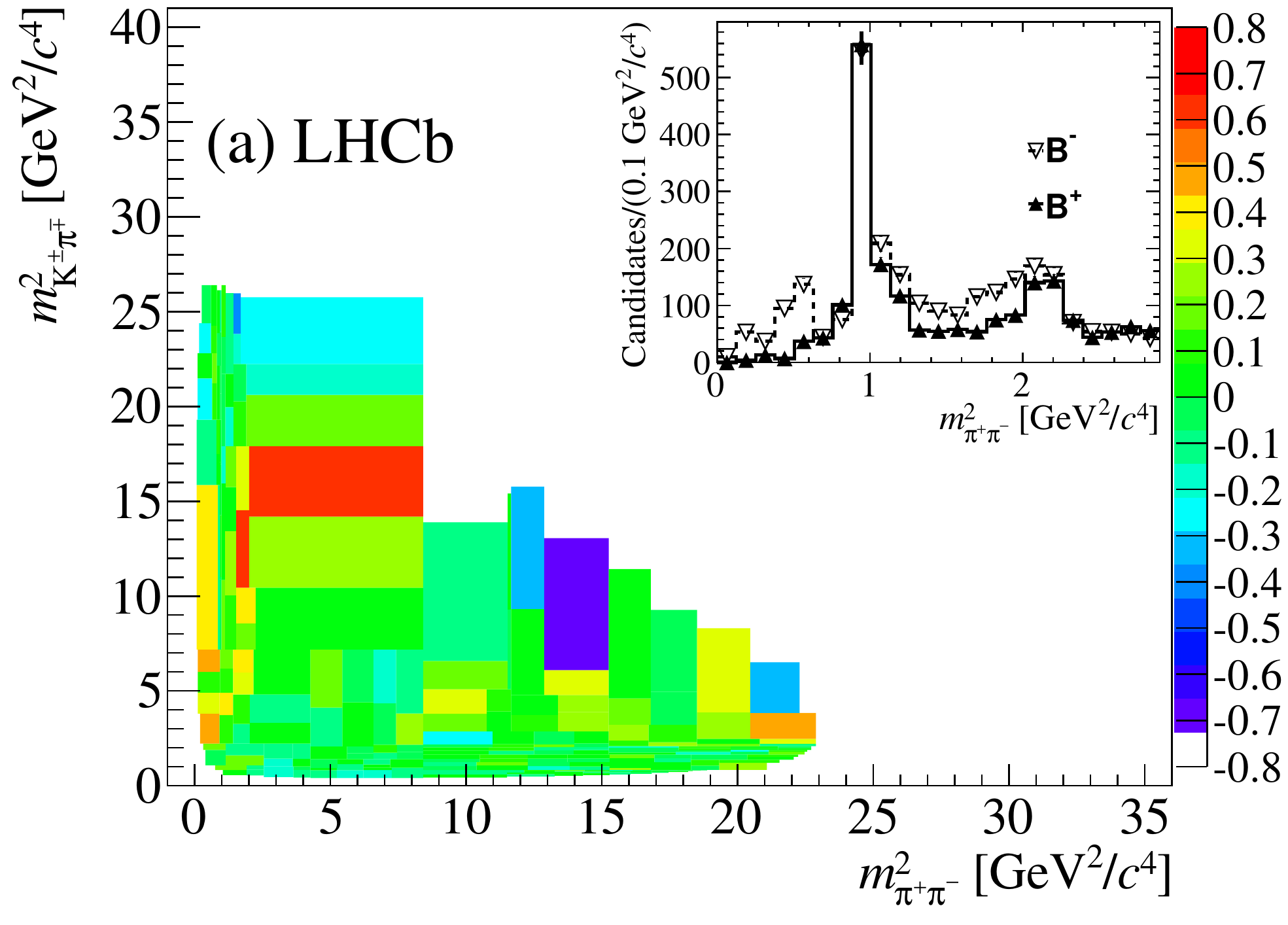}
\includegraphics[width=0.48\linewidth]{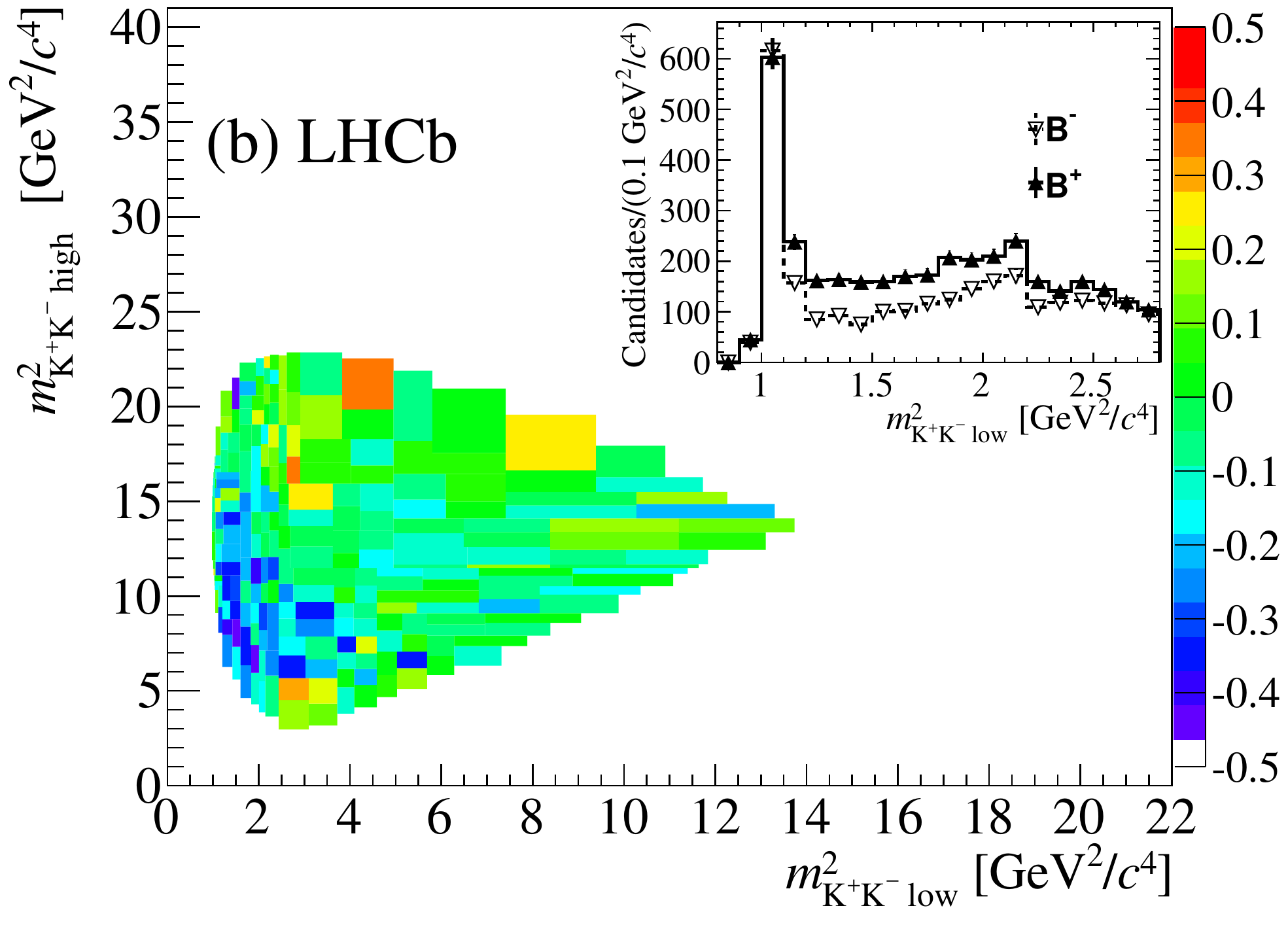}
\caption{Raw asymmetries in bins of the Dalitz plot for (a) \kpipi and (b) \kkk. The inset plots show the projections of the number of background-subtracted events in bins of (a) the \mmpipi variable for $\mmkpi<15\gevgevcccc$ and (b) the \mmkklow variable for $\mmkkhi<15\gevgevcccc$.
}
\label{fig:KpipiDP}
\end{figure}

The asymmetry distributions across the Dalitz planes of these two modes are also investigated. 
The background-subtracted Dalitz plots of the signal region are divided into bins with equal numbers of events, and an asymmetry variable is calculated between the number of negative and positive candidates in each bin, as shown in Fig.~\ref{fig:KpipiDP}. 
The insets show the regions where large local asymmetries are concentrated. 
These appear to be positive for \kpipi, located at low $\pip\pim$ invariant masses below and around the  $\rho(770)^0$ resonance, and above the $f_0(980)$ resonance. 
For \kkk the asymmetries are negative at low values of \mmkklow and \mmkkhi, outside the $\phi(1020)$ resonance and in the region $1.2 < \mmkklow < 2.0\gevgevcccc$ devoid of resonances. 
The local \CP asymmetries are measured from fits to the invariant mass spectra of candidates in two regions where large asymmetries are identified. 
The local charge asymmetries for the regions of $\mmkpi < 15\gevgevcccc$ and $0.08 < \mmpipi < 0.66\gevgevcccc$ for \kpipi decays, and $\mmkkhi < 15\gevgevcccc$ and $1.2 < \mmkklow < 2.0\gevgevcccc$ for \kkk decays, are measured to be
\begin{eqnarray}
 \acp^{\mathrm {reg}}( K\pi\pi)   &=& 0.678 \pm 0.078\stat \pm 0.032\syst \pm 0.007(\jpsiks) ,  \nonumber  \\ [2mm]
\acp^{\mathrm {reg}}(K\!K\!K)   &=&  -0.226  \pm 0.020\stat \pm 0.004\syst \pm 0.007(\jpsiks)  . \nonumber
\end{eqnarray}
The opposite-sign asymmetries between \kpipi and \kkk, concentrated at low $\pip\pim$ and $\Kp\Km$ invariant masses, could be related to compound \CP violation.

\subsection{Comparison of \kkk results}

\begin{figure}[tb]
\centering
\includegraphics[width=0.48\linewidth]{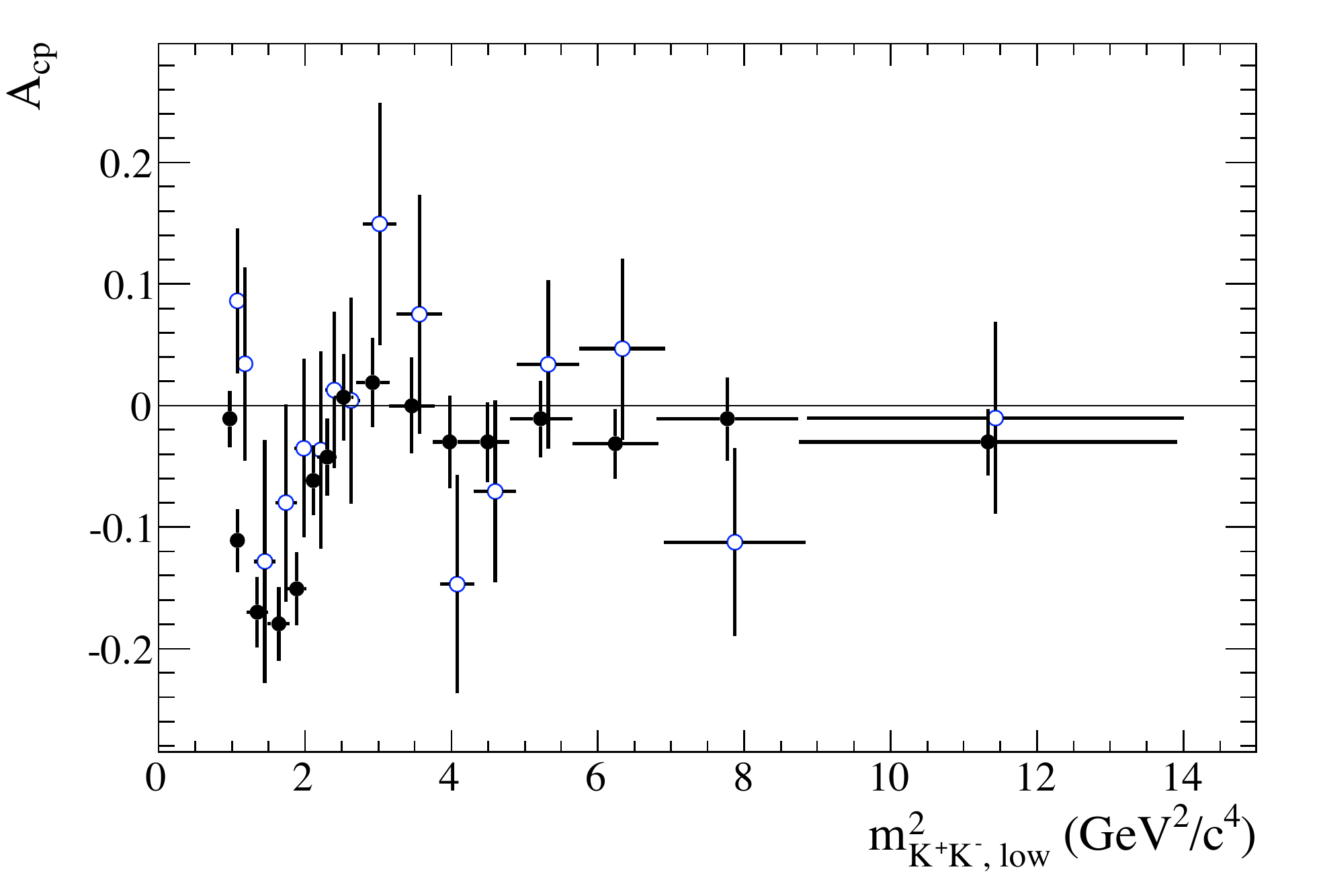}
\includegraphics[width=0.48\linewidth]{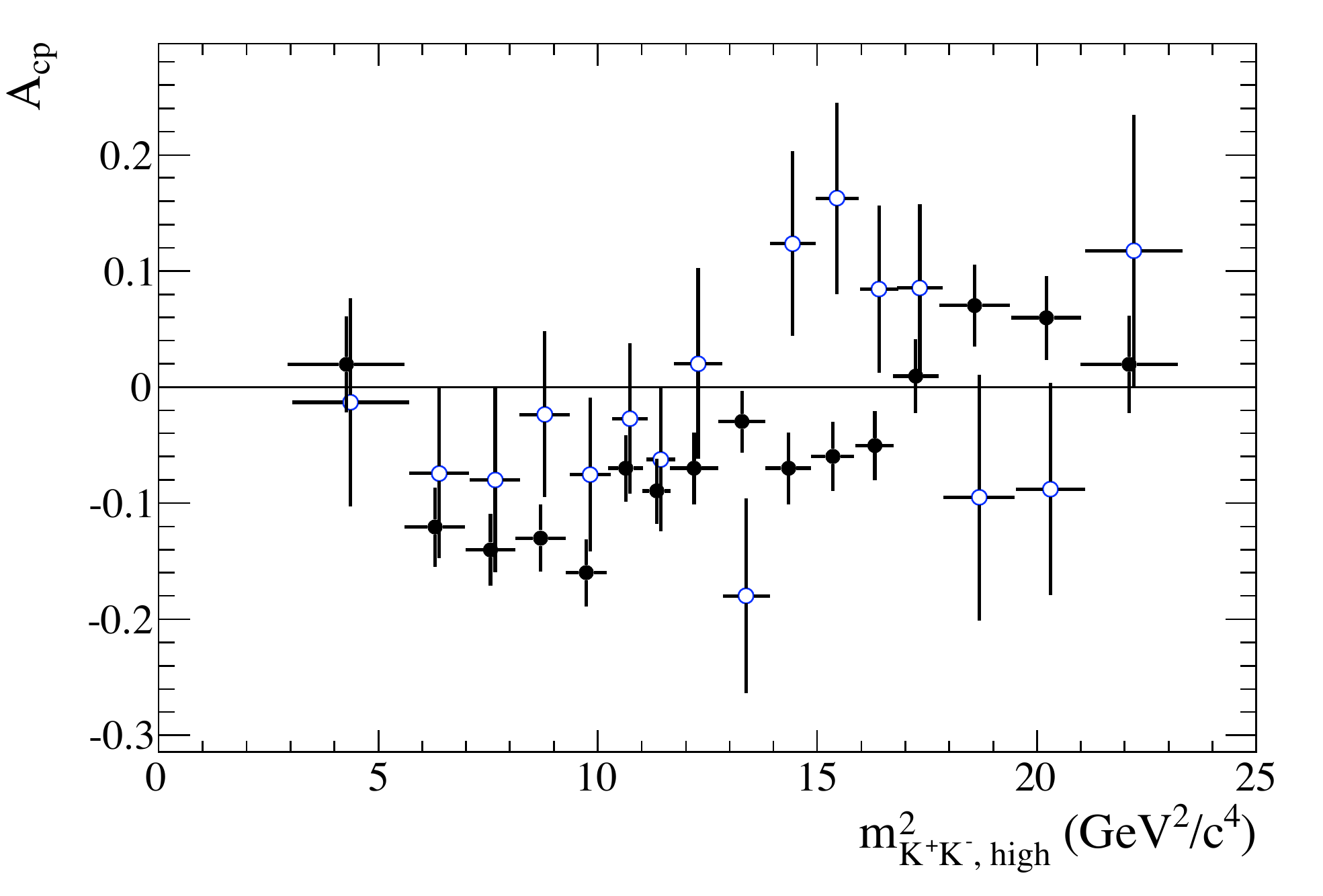}
\caption{Asymmetry of \kkk decays as a function of (left) \mmkklow and (right) \mmkkhi. 
The open points show BaBar measurements obtained using the \splot technique, and the solid points are \acpraw distributions from LHCb.
}
\label{fig:KKKBabar}
\end{figure}

A follow-up study of the amplitude analysis of \kkk decays~\cite{BabarKKK} was performed by BaBar to compare the \CP asymmetry distributions with the preliminary results obtained by LHCb~\cite{LHCb-CONF-2012-018}. 
The study~\cite{BabarKKKnew} investigates the dependence of the \CP asymmetry on the two-body $\Kp\Km$ invariant masses, shown in Fig.~\ref{fig:KKKBabar}. 
The BaBar distributions show \CP asymmetries obtained with the \splot technique, while the LHCb distributions show raw asymmetries from fits to the invariant mass spectra in bins of two-body masses, not corrected for detection and production asymmetries. 
Although the uncertainties on BaBar data are larger than those of LHCb, the patterns of the asymmetries agree well. 
For \mmkklow the asymmetries follow the same shape but seem to have a relative shift, which was observed  to be flat across the spectrum and with an average value of $\Delta({\rm BaBar} - {\rm LHCb})=0.045 \pm 0.021$. 
A similar positive and uniform shift of $0.053\pm 0.021$ was measured from the \mmkkhi distributions.
These shifts are consistent with zero within 2 standard deviations, and are also consistent with the difference between the inclusive \CP asymmetries obtained by BaBar and LHCb.

\subsection{\CP violation in \kkpi and \pipipi decays}

\begin{figure}[tb]
\centering
\includegraphics[width=0.48\linewidth]{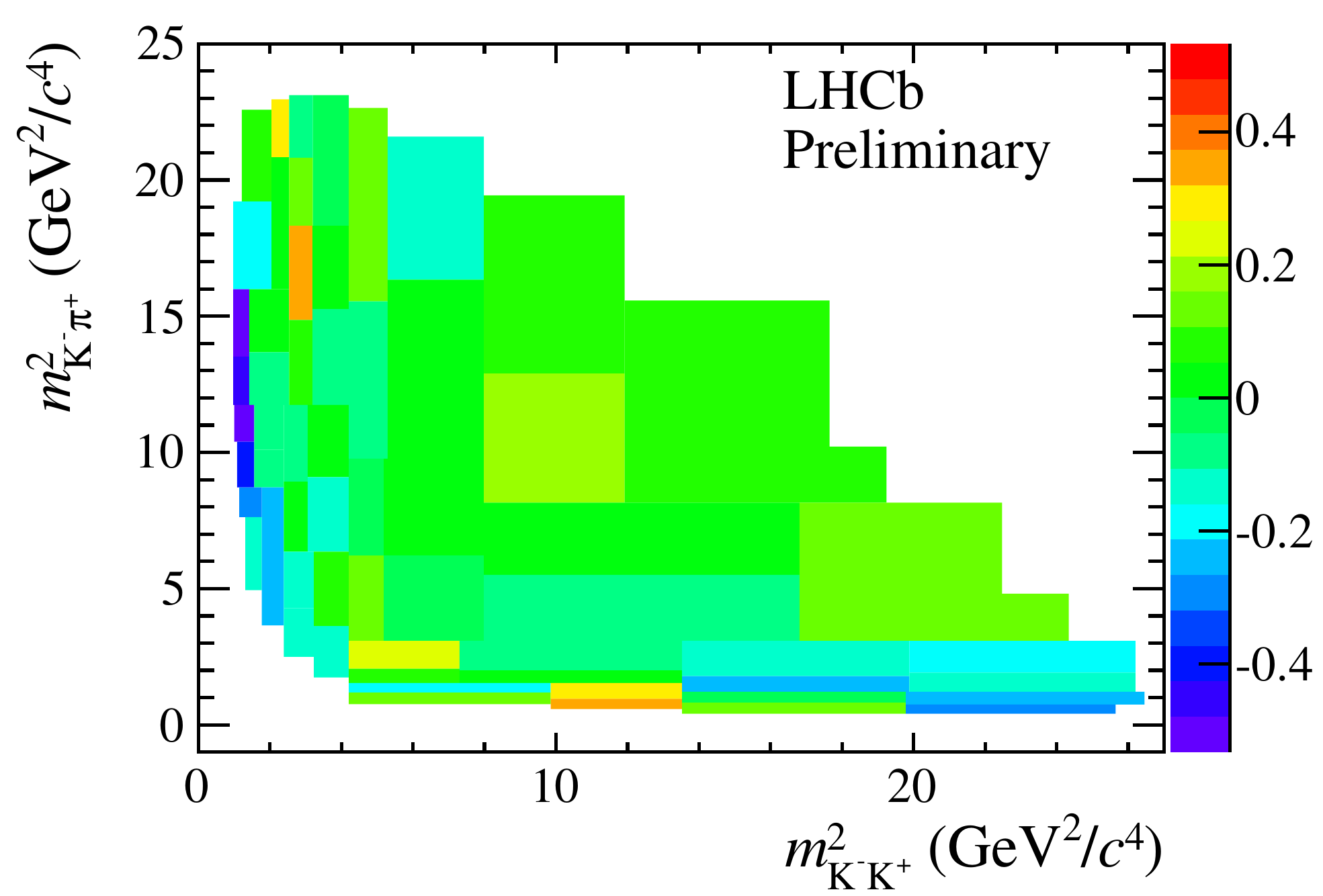}
\includegraphics[width=0.48\linewidth]{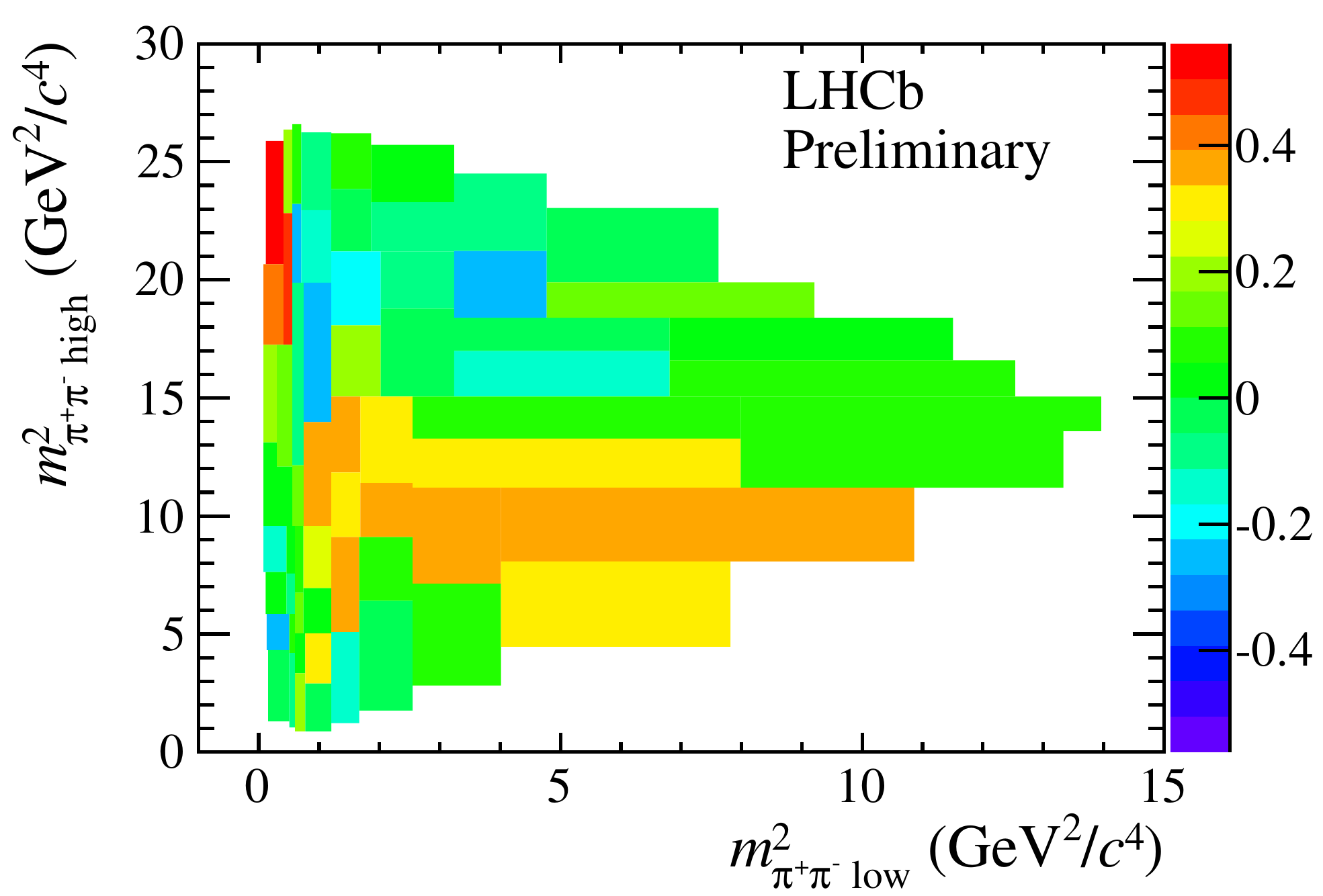}
\caption{Asymmetries on the yields in bins of the Dalitz plane for (left) \kkpi and (right) \pipipi. 
}
\label{fig:pipipi}
\end{figure}

The \kkpi and \pipipi modes are another set of channels with the same flavour quantum numbers, similar to \kpipi and \kkk, thus motivating a joint study. 
A preliminary measurement by LHCb~\cite{LHCb-CONF-2012-028} presents the inclusive \CP asymmetries and their distributions in the Dalitz plots. 
The inclusive charge asymmetries are obtained analogously to the method described in Section~\ref{secKpipi} to be
\begin{eqnarray}
\acp(\pipipi)&=& 0.120 \pm 0.020 \stat \pm  0.019 \syst \pm 0.007(\jpsi\Kpm) \, , \nonumber \\
\acp(\kkpi)&=& -0.153 \pm 0.046 \stat \pm 0.019 \syst  \pm 0.007(\jpsi\Kpm) \, , \nonumber
\end{eqnarray}
with significances of $4.2\sigma$ and $3.0\sigma$, respectively. 
Both results represent the first evidences of inclusive charge asymmetries in these modes. 

The asymmetry distributions of the numbers of events (containing signal and background) in bins of the Dalitz plots of the signal regions are shown in Fig.~\ref{fig:pipipi}. 
For the \kkpi decay mode, a concentration of events, associated with negative asymmetries, is seen at low $\Kp\Km$ invariant masses, and was also studied in the event projections of the two-body invariant mass variables. 
This structure was already reported by BaBar~\cite{BaBarKKpi}, but its asymmetry was not studied. 
The large negative asymmetry seen by LHCb is not associated to resonances, as the \kpipi mode does not have a $\phi(1020)$ contribution. 
For \pipipi decays, a large positive asymmetry is identified at low $\pip\pim$ invariant masses, below the $\rho(770)^0$ resonance. 
The local \CP asymmetries of two regions, chosen as $\mmkk< 1.5\gevgevcccc$  for \kkpi and $\mmpipilow < 0.4\gevgevcccc$ and $\mmpipihi > 15\gevgevcccc$ for \pipipi, are measured to be
\begin{eqnarray}
\acp^{\mathrm {reg}}(\pipipi)   &=&  0.622  \pm 0.075\stat \pm 0.032\syst \pm 0.007(\jpsiks) , \nonumber \\
 \acp^{\mathrm {reg}}( \kkpi)   &=& -0.671 \pm 0.067\stat \pm 0.028\syst \pm 0.007(\jpsiks) . \nonumber   
\end{eqnarray}
The pattern of negative asymmetries at low $\Kp\Km$ masses and positive ones at low $\pip\pim$ masses, not obviously related to resonances, repeats the one seen in \kpipi and \kkk decays. 
These apparent correlations between the pairs of channels could be caused by the \CPT constraint, and further studies need to take this possibility into account.

\subsection{Evidence for the decay \kkpiz}

\begin{figure}[tb]
\centering
\includegraphics[width=0.43\linewidth]{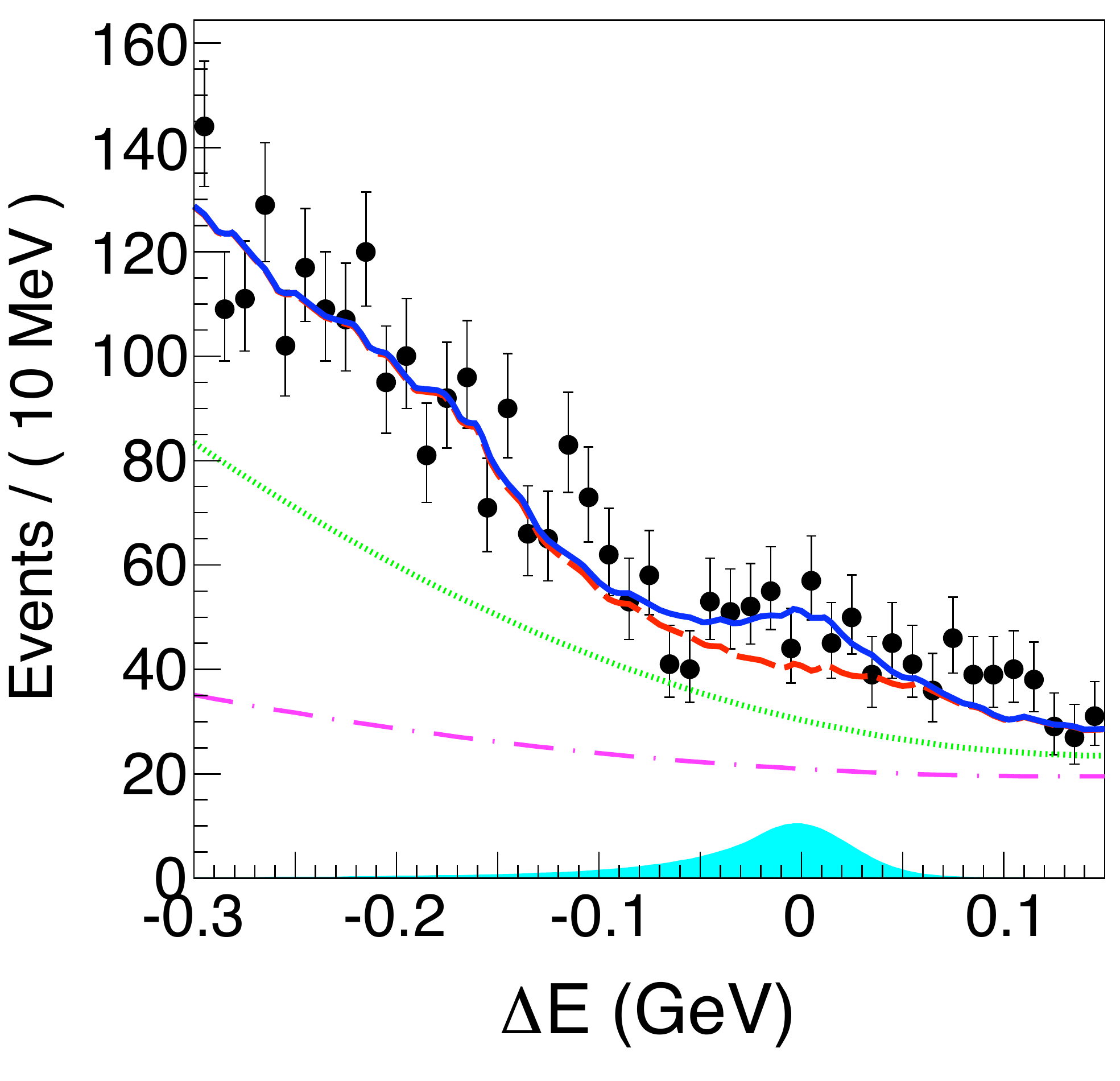}
\includegraphics[width=0.43\linewidth]{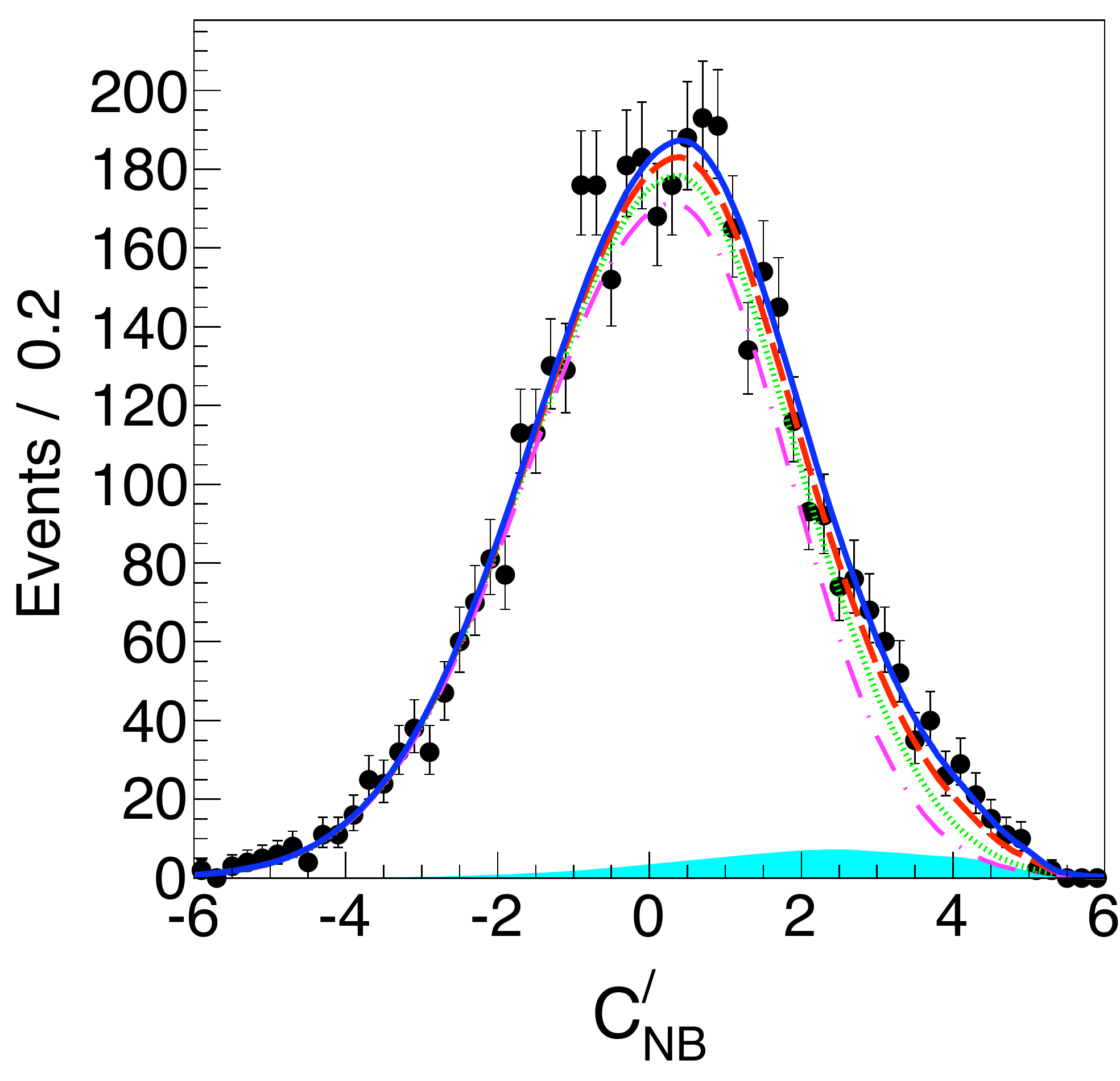}
\caption{Projections of candidate events into (left) $\Delta E$ and (right) $C'_{NB}$. The solid blue curve shows the total PDF, the filled cyan region shows the signal, the dashed red curves the total background, the dotted green curves the continuum $q\bar{q}$ and generic $B\bar{B}$ background, and the dash-dotted magenta curves show the continuum $q\bar{q}$ background.
}
\label{fig:kkpiz}
\end{figure}

A new measurement by Belle of the suppressed \kkpiz decay~\cite{Belle_KKpiz} is presented. 
The data sample is selected using a neural network, and the signal yield is determined from an unbinned extended maximum likelihood fit to the two-dimensional distributions of the energy difference, $\Delta E = \Sigma_i E_i - E_{\rm beam}$, and a variable $C'_{NB}$ related to the neural network output. 
The results of the fit are shown in Fig.~\ref{fig:kkpiz}. 

From a signal yield of $299\pm 83$ events, the branching fraction of the decay is obtained to be
\begin{eqnarray}
{\cal{B}}( \kkpiz) = [2.17 \pm 0.60\stat \pm 0.24\syst]\times 10^{-6} . \nonumber   
\end{eqnarray}
The dominant systematic uncertainties come from the \piz detection efficiency and the efficiency variation over the Dalitz plane. 
The significance of the measurement, determined using a convolution of the statistical likelihood with a Gaussian function of width equal to the systematic errors, is $3.5\sigma$, and constitutes the first evidence for the decay. 

The $\Kp\Km$ and $\Kp\piz$ invariant mass distributions of the signal yields are studied to investigate the resonant structure of the decay. 
The distributions are obtained from fits in bins of the invariant masses. 
An excess of events is observed around 1.4~GeV in the $\Kp\piz$ invariant mass, but its interpretation would require higher statistics. 
The $\Kp\Km$ invariant mass distribution does not allow a definitive statement about a possible structure similar to the one seen in \kkpi decays by BaBar~\cite{BaBarKKpi} and LHCb~\cite{LHCb-CONF-2012-028}.

\subsection{Branching fraction measurements of \ppk decays}

\begin{figure}[tb]
\centering
\includegraphics[width=0.45\linewidth]{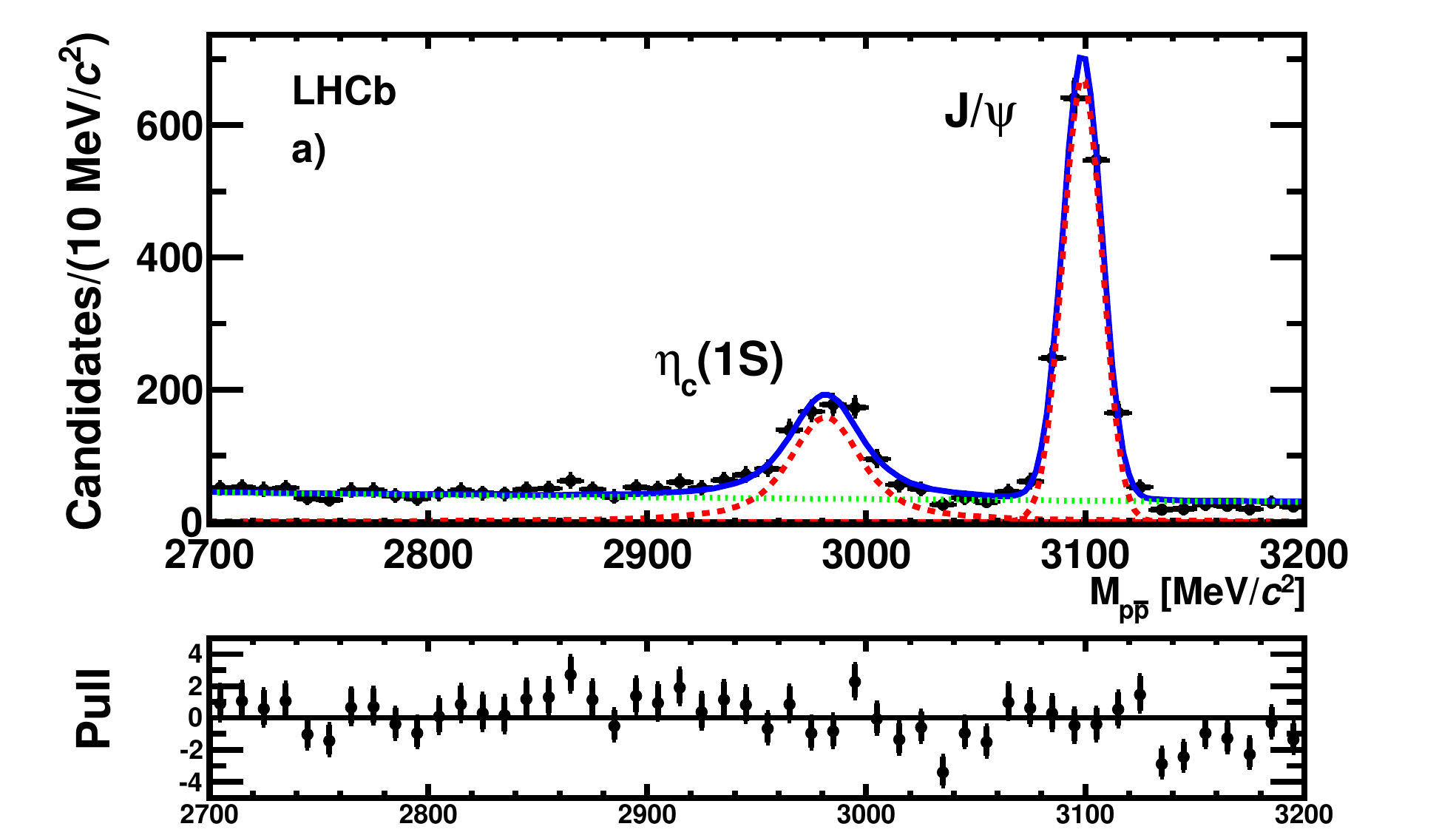}
\includegraphics[width=0.45\linewidth]{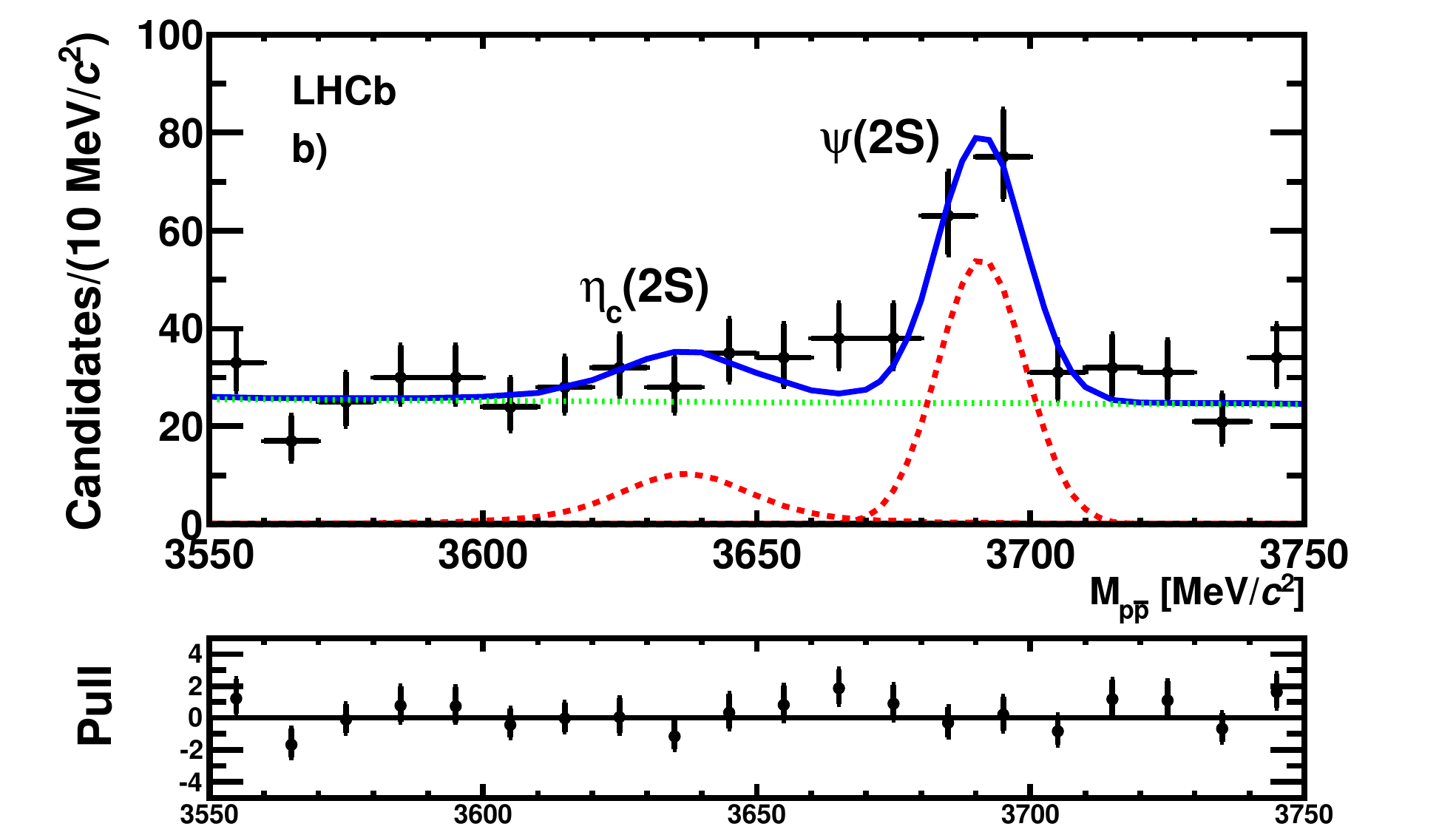}
\caption{Invariant mass distribution of the $p\bar{p}$ system in the regions around (a) the $\eta_c(1S)$ and \jpsi and (b) the $\eta_c(2S)$ and $\psi(2S)$ states.
}
\label{fig:ppk1}
\end{figure}

\begin{figure}[tb]
\centering
\includegraphics[width=0.45\linewidth]{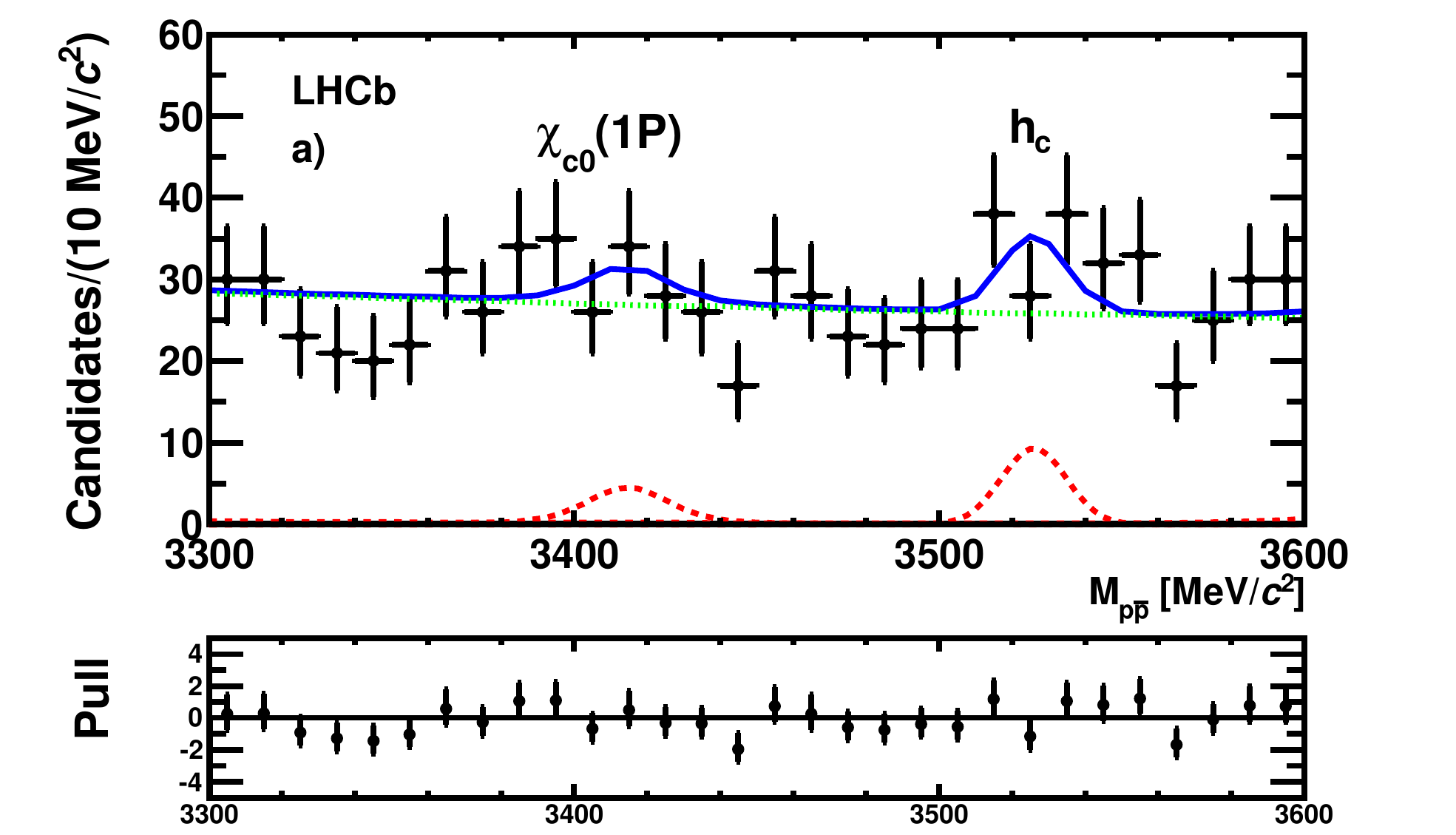}
\includegraphics[width=0.45\linewidth]{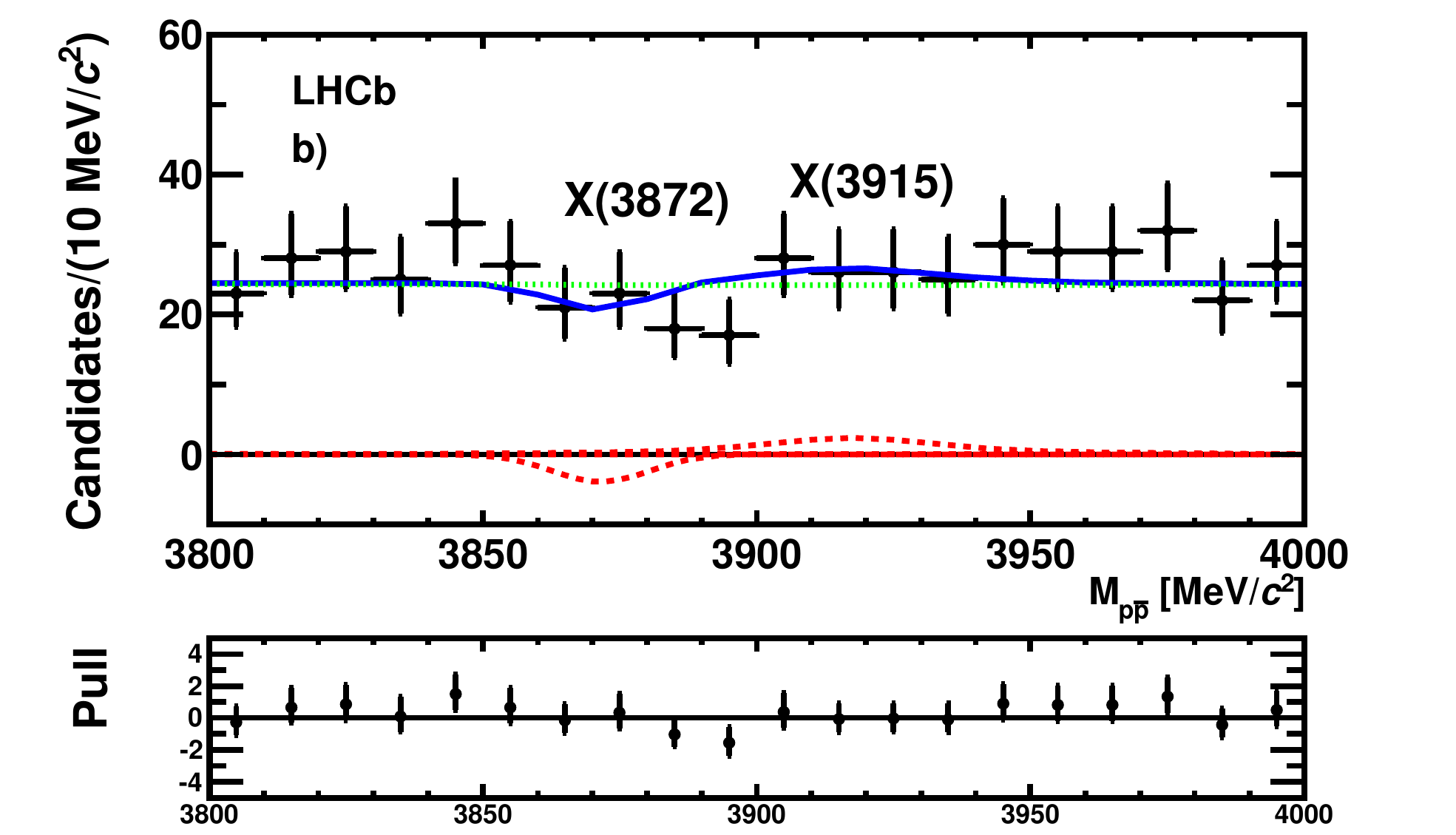}
\caption{Invariant mass distribution of the $p\bar{p}$ system in the regions around (a) the $\chi_{c0}(1P)$ and $h_c$ and (b) the $X(3872)$ and $X(3915)$ states.
}
\label{fig:ppk2}
\end{figure}

The \ppk decay mode allows the study of $c\bar{c}$ states and charmonium-like mesons that decay to $p\bar{p}$. 
The branching fractions of the decay \ppk for different intermediate states have been measured by LHCb~\cite{LHCb-PAPER-2012-047}. 
The event selection is based on a boosted decision tree algorithm. 
The signal yields of the different resonant components are obtained from fits to the $p\bar{p}$ invariant mass distributions within a $B$ mass signal window. 
The fit results are shown in Figs.~\ref{fig:ppk1} and~\ref{fig:ppk2}. 

The branching fractions of the contributions shown in the figures are obtained relative to the one of the \jpsi intemediate state, 
\begin{eqnarray}
\frac{{\cal{B}}(\ppk)_{\rm total}  }{{\cal{B}}(\Bpm \to \jpsi \Kpm \to p\bar{p}\Kpm) }   &=&  4.91  \pm 0.19 \stat \pm 0.14\syst , \nonumber \\
\frac{{\cal{B}}(\ppk)_{M_{p\bar{p}}<2.85\gevcc}  }{{\cal{B}}(\Bpm \to \jpsi \Kpm \to p\bar{p}\Kpm) }   &=&  2.02  \pm 0.10 \stat \pm 0.08\syst , \nonumber \\
\frac{{\cal{B}}(\Bpm \to \eta_c(1S) \Kpm \to p\bar{p}\Kpm)  }{{\cal{B}}(\Bpm \to \jpsi \Kpm \to p\bar{p}\Kpm) }   &=&  0.578  \pm 0.035 \stat \pm 0.025\syst , \nonumber \\
\frac{{\cal{B}}(\Bpm \to \psi(2S)\Kpm \to p\bar{p}\Kpm)  }{{\cal{B}}(\Bpm \to \jpsi \Kpm \to p\bar{p}\Kpm) }   &=&  0.080  \pm 0.012 \stat \pm 0.009\syst . \nonumber 
\end{eqnarray}
The results are compatible with world average values. 
An upper limit on the relative branching fraction of \Bpm into the $X(3872)$ state is also obtained.

\section{Observation of the decay \phik}
\label{BVV}

\begin{figure}[tb]
\centering
\includegraphics[width=0.6\linewidth]{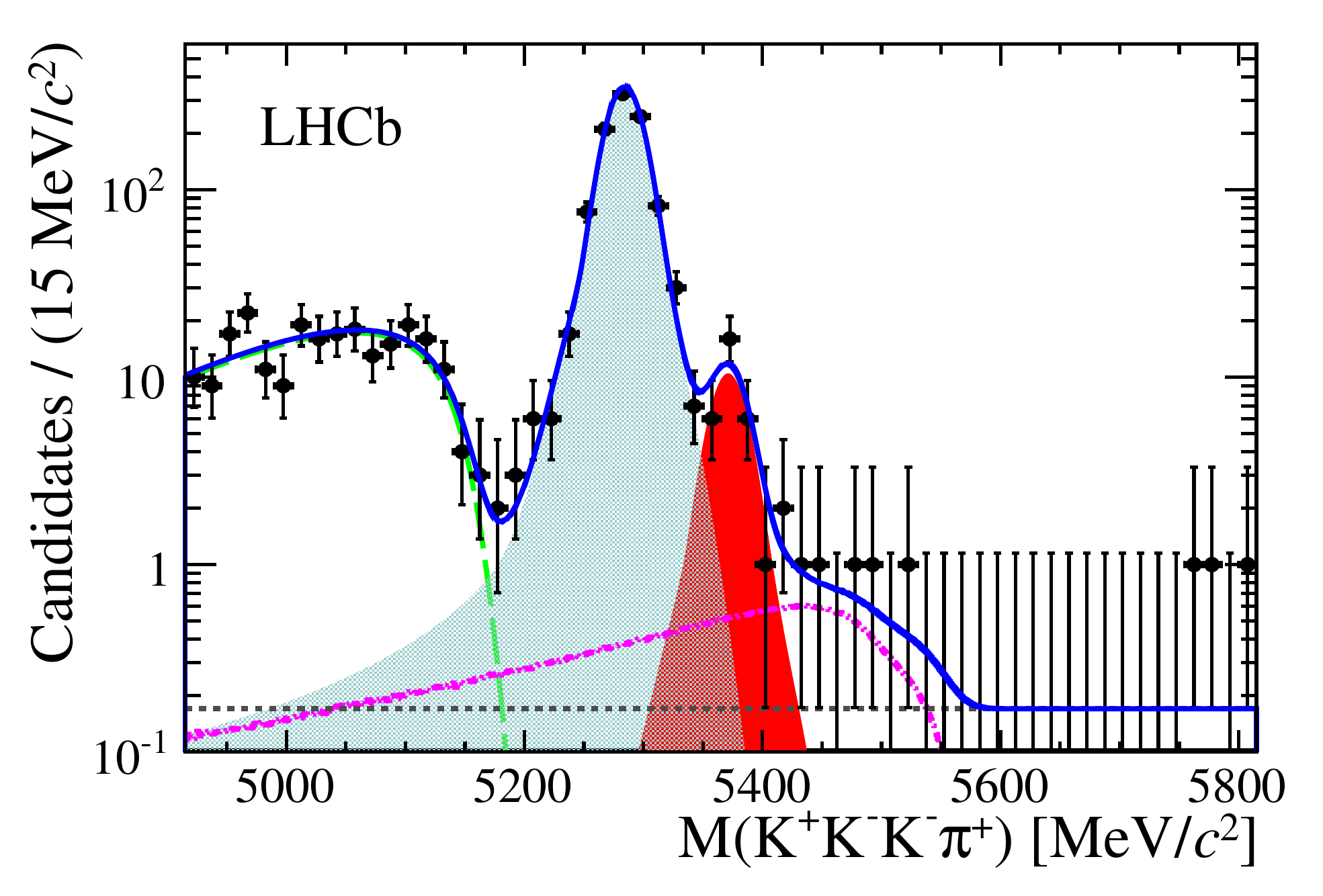}
\caption{Fit to the $KKK\pi$ invariant mass spectrum, with the \phik signal given by the filled red area. 
}
\label{fig:phik}
\end{figure}

$B$ meson decays to two vector mesons that proceed only through penguin diagrams are sensitive to new physics
contributions in the loops. 
They also offer the possibility to study angular distributions and polarisation fractions. 
LHCb has recently claimed the first observation of the \phik decay mode~\cite{LHCb-PAPER-2013-012}. 
The data selection reconstructs the final states  $\Bs \to \phi(\Kp\Km)\Kstarzb(\Km\pip)$ using a geometrical-likelihood multivariate analysis. 
The signal yield is obtained from an unbinned extended maximum likelihood fit to the invariant mass spectrum, shown in Fig.~\ref{fig:phik}, resulting in $30\pm6$ events. 
The statistical significance of the yield is calculated from a likelihood ratio test. 
With a significance of $6.1\sigma$, this measurement represents the first observation of the \phik decay mode. 

The branching fraction of the decay is measured using $\Bd\to\phi\Kstarz$ as a normalisation channel to be
\begin{eqnarray}
{\cal{B}}(\phik)  =  [1.10  \pm 0.24 \stat \pm 0.14\syst \pm 0.08 (f_d/f_s)]\times 10^{-6} , \nonumber 
\end{eqnarray}
where the third uncertainty is due to the uncertainty on the ratio of $b$-hadron production fractions $f_d/f_s$~\cite{fdfs}. 
The main systematic uncertainties are due to the fit model and the purity of the resonance components. 

An untagged time-integrated polarisation analysis of the angular distributions is performed to obtain the polarisation fractions of the decay. The longitudinal polarisation fraction is measured to be
\begin{eqnarray}
f_0  =  0.51  \pm 0.15 \stat \pm 0.07\syst  . \nonumber 
\end{eqnarray}
The result agrees with that of the $b \to s$ penguin decay $\Bd\to\phi\Kstarz$~\cite{polar}.

\section{Conclusions}

Several recent results from studies of charmless $B$ decays were reported. 
Branching fraction measurements include the first observation of \phik decays and the first evidence for \kkpiz decays. 
The first observation of direct \CP violation in \Bs decays was presented, as well as evidences for inclusive charge asymmetries in three-body \Bpm decays.  
Interesting patterns of \CP violation in three-body charmless \Bpm decays were observed by LHCb. 
Further experimental studies are needed to resolve the difference between the \CP asymmetries of \kkk seen by BaBar and LHCb.

\Acknowledgements
I am grateful to the CBPF and the organising committee of FPCP--2013 for funding my participation in the conference.

\end{document}